\documentclass[]{elsarticle}

\usepackage{lineno,hyperref}
\modulolinenumbers[5]

\journal{Journal of \LaTeX\ Templates}

\begin{document}

\begin{frontmatter}

\title{On the Filamentation Instability in Degenerate Relativistic Plasmas}
\tnotetext[mytitlenote]{Fully documented templates are available in the elsarticle package on \href{http://www.ctan.org/tex-archive/macros/latex/contrib/elsarticle}{CTAN}.}


\author{Goshadze R.M.$^{1}$, Berezhiani V.I.$^{1,2}$ and Osmanov Z.$^{1}$}

\address{$^{1}$School of Physics, Free university of Tbilisi, Tbilisi 0159, Georgia}
\address{$^{2}$Andronikashvili Institute of Physics (TSU), Tbilisi \ 0177, Georgia.}






\begin{abstract}
The filamentation instability of the electromagnetic (EM) beam in an underdense plasma with high level of degeneracy is examined by means of the momentum equation, continuity
equation and Maxwell's equations. It has been demonstrated that the instability develops for weakly as
well as strongly relativistic degenerate plasma and arbitrary strong amplitude
of EM beams.\end{abstract}

\begin{keyword}
Degenerate relativistic plasma; Filamentation; Instability
\end{keyword}

\end{frontmatter}

\section{Introduction}

From observations it is evident that certain class of astronomical sources have 
extremely high luminosities covering
almost the whole range of electromagnetic spectra 
\cite{Begelman}. Usually, for the internal structures of post-main sequence stars, with dense magnetospheric plasma
\cite{Shapiro}, the convenient approach is not working any more because of the high degeneracy of electrons. If this is the case, the physical system has to be described by the Fermi-Dirac distribution
when the corresponding Fermi energy, $\epsilon_{F}$ exceeds 
that of the binding energy of electrons, which leads to ionisation of atoms. 
For such a system the concentration of elctrons is
 of the order of $10^{26-34}cm^{-3}$ and consequently the mean spacing between particles
 is small compared to the thermal 
de Broglie wavelength. \cite{Landau}.

One can straightforwardly show that for electron concentrations, $n$,
exceeding the critical value $n_{c}=m_{e}^{3}c^{3}/3\pi
^{2}\hbar^{3}=5.9\times10^{29}cm^{-3}$, the energy 
of Fermi level $\epsilon_{F}=m_{e}c^{2}\left(  \gamma_{F}-1\right)$, might exceed the rest mass energy,
implying that at such densities, the degenerate electron gas must be treated relativistically even if its
"temperature" is nonrelativistic, or even zero. Here,
$\gamma_{F}=\sqrt{1+p_{F}^{2}/m_{e}^{2}c^{2}}$ and $p_{F}=m_{e}c\left(
n/n_{c}\right)  ^{1/3}$ represent the Fermi relativistic factor and relativistic momentum respectively \cite{Haas}.

In a series of papers \cite{Guga}-\cite{Nana}, where authors examined the nonlinear
character of interactions of plasma waves and high frequency EM 
waves it was found that regardless of the degeneracy stable localised EM structures are induced.
The stimulated Raman scattering (SRS) instability for such a
plasma was studied in one dimensional case
\cite{Gio-Raman}. The authors have shown that the degenerate relativistic plasma 
reveals interesting properties of SRS instability in low density plasma with the frequency
$\omega>2\omega_{e}$,
where $\omega_{e}=\left(
4\pi e^{2}n_{_{0}}/m_{e}\right)  ^{1/2}$ represents the plasma frequency and $n_{0}$
is equilibrium number density of electrons. For highly dense plasma, the radiation process
might potentially lead to hard X-rays, with specific observational features.

Intense EM waves may undergo filamentation instability
(FI) in the underdense plasma. This process inevtably leads to 
break up of the EM field into multiple beamlets in the
direction perpendicular to the incoming radiation \cite{Tsigni}. The linear as
well as nonlinear regime of FI has been actively
studied in nonrelativistic and relativistic intense EM
beams \cite{Litvak}-\cite{Sharma}. However, to the best of our knowledge, the
FI of EM radiation in degenerate relativistic plasma is not
addressed so far.

In the present paper, we apply the fluid-Maxwell model developed in
\cite{Guga},\cite{Nana1} to study the possibility of FI of intense
narrow electromagnetic pulse $L_{\perp}<<L_{\parallel}$ (where $L_{\parallel}$ and
$L_{\perp}$ are the characteristic longitudinal and transverse spatial
dimensions of the field, respectively) in the transparent degenerate electron plasma 
to show the possibility of FI in relativistic degenerate plasma embedded in the field of arbitrary strong
EM radiation.

\section{Main Consideration}
Our approach is based on methods and tools developed in previous works. In particular,
we imply the Maxwell's equations and the
relativistic electron plasma fluid model. Throughout the paper it is assumed that
the thermal energy of electrons
is negligible compared to their Fermi energy. In the framework of the model
the ions are considered to be in stationary states, forming neutralizing background. For
zero generalized vorticity $\mathbf{\Omega=\nabla\times}\left(  G\mathbf{p-}%
e\mathbf{A/}c\right)  =0$, the electron fluid equations are given by (see Ref. \cite{Guga} for details):

\begin{equation}
\frac{\partial}{\partial t}\left(  G\mathbf{p-}e\mathbf{A}/c\right)
+\mathbf{\nabla}\left(  m_{c}c^{2}G\gamma-e\varphi\right)  =0, \label{R1}%
\end{equation}

\begin{equation}
\frac{\partial}{\partial t}N+\nabla\cdot\left(  N\mathbf{V}\right)  =0.
\label{R2}%
\end{equation}
The corresponding Maxwell's equations in the Coulomb gauge $\mathbf{\nabla\cdot A}=0$
are expressed as follows :%
\begin{equation}
\frac{\partial^{2}\mathbf{A}}{\partial t^{2}}-c^{2}\Delta\mathbf{A+}%
c\frac{\partial}{\partial t}\left(  \mathbf{\nabla}\varphi\right)  +4\pi
ecN\mathbf{V}=0 \label{R3}%
\end{equation}

\begin{equation}
\Delta\varphi=4\pi e(N-n_{0}). \label{R4}%
\end{equation}
where $\mathbf{A}$ and $\varphi$ represent the EM field vector and scalar
potentials respectively;\textbf{ }$\mathbf{p=}m_e\gamma\mathbf{V}$ denotes the hydrodynamic
momentum of electrons, $\mathbf{V}$ is the velocity and $\gamma=\left(
1+p^{2}/m_{e}^{2}c^{2}\right)  ^{1/2}$ is the Lorentz factor; $N$ denotes the
electron number density in the laboratory frame of reference, that is related to the rest
frame electron density by following $N=\gamma n$. It is evident from Eq.(\ref{R1})
that for the arbitrary strength of relativity defined by the ratio $n/n_{c}$
the \textquotedblleft effective mass\textquotedblright\ factor $G$ for fully
degenerate plasma (i.e. having zero temperature) coincides with the Fermi
relativistic factor $G=\gamma_{F}=\left(  1+\left(  n/n_{c}\right)
^{2/3}\right)  ^{1/2}$.

To study the problem of the nonlinear
self-guiding of the EM beam in a highly transparent electron plasma we apply
Eqs. (\ref{R1}-\ref{R4}), which for the generalized momentum $\mathbf{\Pi=}G\mathbf{p}$
and relativistic factor $\Gamma=G\gamma$ reduce to the following set of dimensionless equations

\begin{equation}
\frac{\partial}{\partial t}\left(  \mathbf{\Pi-A}\right)  +\mathbf{\nabla
}\left(  \Gamma-\varphi\right)  =0, \label{R5}%
\end{equation}

\begin{equation}
\frac{\partial}{\partial t}N+\nabla\cdot\mathbf{J}=0, \label{R6}%
\end{equation}

\begin{equation}
\frac{\partial^{2}\mathbf{A}}{\partial t^{2}}-\Delta\mathbf{A+}\frac{\partial
}{\partial t}\left(  \mathbf{\nabla}\varphi\right)  +\varepsilon^{2}%
\mathbf{J}=0, \label{R7}%
\end{equation}

\begin{equation}
\Delta\varphi=N-1, \label{R8}%
\end{equation}
were $\mathbf{J=}N\mathbf{\Pi}/\Gamma$ and $\Gamma=\left(  G^{2}+\mathbf{\Pi
}^{2}\right)  ^{1/2}$, $\widetilde{t}=\omega t$, $\widetilde{\mathbf{r}}=\omega
\mathbf{r/}c$, $\widetilde{\mathbf{A}}=e\mathbf{A}/m_{e}c^{2}$, $\widetilde
{\varphi}=e\varphi/m_{e}c^{2}$, $\widetilde{\mathbf{\Pi}}=\mathbf{\Pi}/m_{e}%
c$, $\widetilde{n}=n/n_{0}$ and $\widetilde{N}=N/n_{0}$ (in the above equations the tilde
is omitted). Here we assume that the plasma is highly underdens, $\varepsilon=\omega_{e}/\omega<<1$, where $\varepsilon$ is
a small parameter of the system and $G=\left(  1+R_{0}^{2}n^{2/3}\right)
^{1/2}$ where $R_{0}=\left(  n_{0}/n_{c}\right)  ^{1/3}$. 

Following the method of multiple scale expansion of the equations in the small parameter
$\varepsilon$ \cite{Sun}, \cite{Ohashi}, the physical variables ($Q=A,\varphi
,\Pi,\Gamma,N,G$) expand as%

\begin{equation}
Q=Q_{\left\{  0\right\}  }\left(  \xi,x_{1},y_{1},z_{2}\right)  +\varepsilon
Q_{\left\{  1\right\}  }\left(  \xi,x_{1},y_{1},z_{2}\right),  \label{R9}%
\end{equation}
where $\left(  x_{1},y_{1},z_{2}\right)  =\left(  \varepsilon x,\varepsilon
y,\varepsilon^{2}z\right)  $ and $\xi=z-bt$ and $\left(  b^{2}-1\right)
\sim\varepsilon^{2}$. In the framework of the paper we assume that
the EM field is circularly polarized%

\begin{equation}
\mathbf{A}_{\left\{  0\perp\right\}  }=\frac{1}{2}\left(  \widehat{\mathbf{x}%
}+i\widehat{\mathbf{y}}\right)  A\exp\left(  i\xi/b\right),  \label{R10}%
\end{equation}
with a slowly varying function $A$.

In the zeroth order approximation for the longitudinal (to the direction of EM wave propagation $z$) component of Eq.( \ref{R5}) one can straightforwardly show $\Pi_{\{z0\}}=0$ whereas for the transverse 
component we obtain $\mathbf{\Pi}_{\left\{  0\perp\right\}  }=-\mathbf{A}_{\left\{  0\perp
\right\}  }$. From the transverse part of Eq. (\ref{R5}) to
the first order in $\varepsilon$ and Eq. (\ref{R6}) one can conclude that
$\Gamma_{\{0\}},\varphi_{\{0\}}$ and $N_{\{0\}}$ are independent from the
fast variable $\xi.$ For the transverse part of the first order of Eq. (\ref{R5})
we arrive at
\begin{equation}
-b\frac{\partial}{\partial\xi}\left[  \mathbf{\Pi}_{\perp\left\{  1\right\}
}-\mathbf{A}_{\perp\left\{  1\right\}  }\right]  +\mathbf{\nabla}_{\perp
}\left(  \Gamma_{\left\{  0\right\}  }-\varphi_{\left\{  0\right\}  }\right)
=0, \label{R10a}%
\end{equation}
where $\mathbf{\nabla}_{\perp}=\left(  \widehat{\mathbf{x}}\partial/\partial
x_{1}+\widehat{\mathbf{y}}\partial/\partial y_{1}\right)  $. By averaging
Eq. (\ref{R8}) over the fast variable $\xi$ the equation reduces to

\begin{equation}
\mathbf{\nabla}_{\perp}\Gamma_{\left\{  0\right\}  }=\mathbf{\nabla}_{\perp
}\varphi_{\left\{  0\right\}  } \label{R11},%
\end{equation}
where $\Gamma_{\left\{  0\right\}  }=\left(  G_{\left\{  0\right\}  }%
^{2}+\left\vert A\right\vert ^{2}\right)  ^{1/2}$, $G_{\left\{  0\right\}
}=\left(  1+R_{0}^{2}n_{\left\{  0\right\}  }^{2/3}\right)  ^{1/2}$ and
the rest frame electron density is denoted by $n_{\left\{  0\right\}  }$. This value is related to the
lab frame density by the following relation%

\begin{equation}
n_{\left\{  0\right\}  }=\frac{G_{\left\{  0\right\}  }N_{\left\{  0\right\}
}}{\Gamma_{\left\{  0\right\}  }}. \label{R12}%
\end{equation}
 Eq. (\ref{R11}) leads to $\Gamma_{\left\{  0\right\}
}-\varphi_{\left\{  0\right\}  }=\Gamma_{0}$ provided that EM field intensity forms a constant background $\left\vert
A\right\vert ^{2}\rightarrow\left\vert A_{0}\right\vert ^{2}$ for $\left\vert
r_{\perp}\right\vert \rightarrow\infty$ while $\varphi_{\left\{  0\right\}
}\rightarrow0$ and $N_{\left\{  0\right\}  }\rightarrow1$. After combining this algebraic relation with
Eq. (\ref{R12}) one can obtain the expression for the plasma electron density:%

\begin{equation}
N_{\left\{  0\right\}  }=\left(  \frac{\Gamma_{0}}{R_{0}}\right)  ^{3}%
\frac{\left(  1+\Psi\right)  \left[  \left(  1+\Psi\right)  ^{2}-\left(
1+\left\vert A\right\vert ^{2}\right)  /\Gamma_{0}^{2}\right]  ^{3/2}}{\left[
\left(  1+\Psi\right)  ^{2}-\left\vert A\right\vert ^{2}/\Gamma_{0}%
^{2}\right]  ^{1/2}},\label{R13}%
\end{equation}
where $\Psi$ denotes the normalized value of $\varphi$: $\Psi=\varphi/\Gamma_{0}$.

An "effective" relativistic factor, $\Gamma_{0}$, for the given boundary conditions depends on the
parameters $R_{0}$ and $\ \left\vert A_{0}\right\vert ^{2}$ by the following
implicit expression:%

\begin{equation}
\left(  1-\frac{1+\left\vert A_{0}\right\vert ^{2}}{\Gamma_{0}^{2}}\right)
-\frac{R_{0}^{2}}{\Gamma_{0}^{2}}\left(  1-\frac{\left\vert A_{0}\right\vert
^{2}}{\Gamma_{0}^{2}}\right)  ^{1/3}=0. \label{R14}%
\end{equation}
One can see that when Eq. (\ref{R14}) is satisfied $N_{\left\{  0\right\}  }=1$ for $\Psi->0$.

For the weakly degenerate plasma $R_{0}<<1$ the constant $\Gamma_{0}$ coincides
with the Lorentz factor of electrons in classical cold plasmas
$\Gamma_{0}\simeq\left(  1+\left\vert A_{0}\right\vert ^{2}\right)  ^{1/2}$.
\ In the weakly relativistic amplitudes of EM field ($\left\vert
A_{0}\right\vert \rightarrow0$) $\Gamma_{0}$ is determined by the degeneracy
parameter $\Gamma_{0}\simeq\left(  1+R_{0}^{2}\right)  ^{1/2}$. \ In Fig.1
we show the dependence of $\ \Gamma_{0}$ on $\left\vert A_{0}\right\vert $
for different values of $R_{0}$. In the extreme relativistic amplitudes ($\left\vert A_{0}\right\vert \gg1$),
$\Gamma_{0}$ tends to $\left\vert A_{0}\right\vert $ (essentially the
expression of a nondegenerate plasma) for an arbitrary level of the degeneracy
parameter $R_{0}$.

\begin{figure}
\resizebox{1\textwidth}{!}{%
  \includegraphics{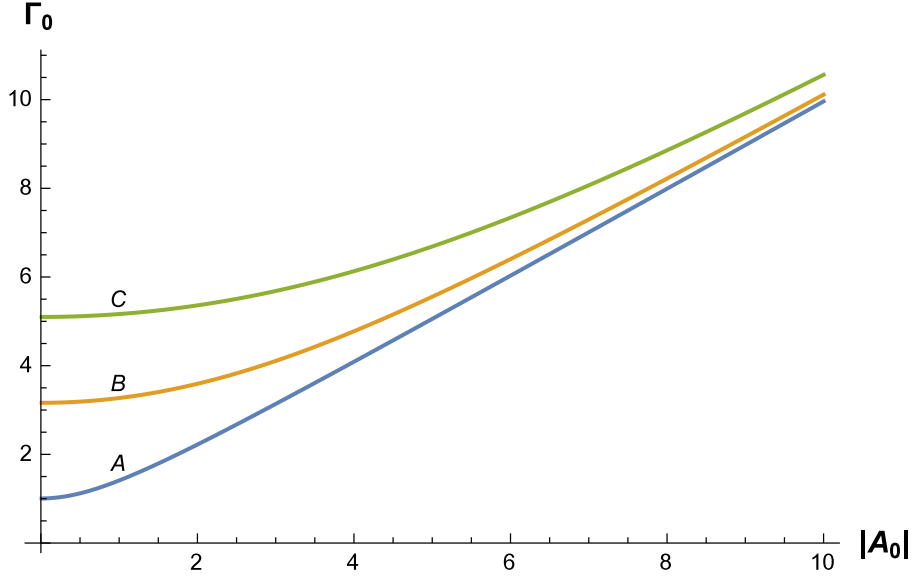}
  }
\caption{Dependence of $\Gamma_{0}$ on the EM field strength $A_{0}$ for
		different level of degeneracy: (A) $R_{0}=0.1$, ( B) $R_{0}=3$, and (C)
		$R_{0}=5$.}
\label{fig1}       
\end{figure}

By considering the slowly varying envelope, Maxwell's equation (\ref{R7}) and the Poisson's equation (\ref{R8}) reduce to%
\begin{equation}
2i\frac{\partial A}{\partial z_{2}}+\nabla_{\perp}^{2}A+\sigma A-A\frac
{N_{\left\{  0\right\}  }}{\Gamma_{\left\{  0\right\}  }}=0, \label{R15}%
\end{equation}

\begin{equation}
\nabla_{\perp}^{2}\Psi=\frac{1}{\Gamma_{0}}\left(  N_{\left\{  0\right\}
}-1\right)  , \label{R16}%
\end{equation}
where $\sigma=\left(  b^{2}-1\right)  /b^{2}\varepsilon^{2}$. For  
$b=\omega/kc$ one can see that $\sigma-1/\Gamma_{0}=0$ leads to 
the following dispersion
relation $\omega^{2}=k^{2}c^{2}+\Omega_{e}^{2}$, with $\Omega_{e}=\omega
_{e}/\Gamma_{0}$ representing a plasma frequency modified by the effective
relativistic factor $\Gamma_{0}$. Introducing the following re-normalizations
$z=z/\Gamma_{0}$ and $r_{\perp}=r_{\perp
}/\Gamma_{0}^{1/2}$ Eqs. (\ref{R15})-(\ref{R16}) can be written as

\begin{equation}
2i\frac{\partial A}{\partial z}+\nabla_{\perp}^{2}A+\left(  1-\frac{N}{\Psi
+1}\right)  A\mathcal{=}0, \label{R17}%
\end{equation}

\begin{equation}
\nabla_{\perp}^{2}\Psi+1-N=0, \label{R18}%
\end{equation}
where we have omitted subscripts for the variables
$\left(  z_{2},x_{1},y_{1},N_{\left\{  0\right\}  }\right)  $.

To study the stability of the ground state solution we linearized the
system of equations (\ref{R17})-(\ref{R18}) in the form $A=A_{0}+\delta A$,
$\Psi=\delta\Psi$, $N=1+\delta N$. By assuming that all perturbed quantities $\left(  \delta A,\delta\Psi,\delta N\right)$ depend
on coordinates as $\exp(\chi
z+i\mathbf{k}_{\perp}\mathbf{r}_{\perp})$, the following dispersion relation can be obtained

\begin{equation}
\chi=\frac{1}{2}k_{\perp}\left(  \frac{1+k_{\perp}^{2}}{\left(  k_{\perp}%
^{2}+C_{\Psi}\right)  }\left\vert A_{0}\right\vert ^{2}C_{a}-k_{\perp}%
^{2}\right)  ^{1/2}\label{R19}%
\end{equation}
that implies the necessary condition for the instability \
\begin{equation}
\frac{1+k_{\perp}^{2}}{\left(  k_{\perp}^{2}+C_{\Psi}\right)  }\left\vert
A_{0}\right\vert ^{2}C_{a}>k_{\perp}^{2},\label{R20}%
\end{equation}
where $C_{a}=-2\left[  \partial N/\partial\left\vert A\right\vert ^{2}\right]
_{\left\vert A\right\vert =\left\vert A_{0}\right\vert ,\Psi=0}$ and $C_{\Psi
}=\left[  \partial N/\partial\Psi\right]  _{\left\vert A\right\vert
=\left\vert A_{0}\right\vert ,\Psi=0}$. \ From Eq. (\ref{R14})
one can show that variables $C_{a}$ , $C_{\Psi}$ are both positive and they can be presented
as $C_{a}=Q/R_{0}^{2}$ and $C_{\Psi}=1+\Gamma_{0}^{2}Q/R_{0}^{2}$ where%

\begin{equation}
Q=2\Gamma_{0}^{-2}\left(  1-\left\vert A_{0}\right\vert ^{2}/\Gamma_{0}%
^{2}\right)  ^{-4/3}\left[  3/2+\Gamma_{0}^{2}-(1+\left\vert A_{0}\right\vert
^{2})\right]. \label{R21}%
\end{equation}

From the structure of \ Eqs. (\ref{R19})- (\ref{R21}) it is evident that
the perturbations with the transverse wave number $0<k_{\perp}<k_{\lim}$, where
$k_{\lim}$ is a certain limiting value, are unstable with respect to $z$. Since Eq.(\ref{R19}) contains several parameters $\left\vert A_{0}\right\vert ,\Gamma_{0}$ and $R_{0}$, which are interrelated
by Eq. (\ref{R14}), in general, should be obtained numerically. However, for small amplitude pump waves
$\left\vert A_{0}\right\vert <<1$, $Q\simeq\left(  1+2\Gamma_{0}^{2}\right)
/\Gamma_{0}^{2}$ the dispersion relation reads as

\begin{equation}
\chi=\frac{1}{2}k_{\perp}\left(  \left\vert A_{0}\right\vert ^{2}K_{\Gamma
_{0}}^{2}-k_{\perp}^{2}\right)  ^{1/2},\label{R22}%
\end{equation}
where $K_{\Gamma_{0}}=\left[  \left(  1+2\Gamma_{0}^{2}\right)  /3\Gamma
_{0}^{4}\right]  ^{1/2}$ and the Lorentz factor determined
by the degeneracy parameter $R_{0}$ is given by $\Gamma_{0}=\left(
1+R_{0}^{2}\right)  ^{1/2}$ (see Eq. (\ref{R14})). For weakly degenerate
plasma ($R_{0}<<1$) \ \ $K_{\Gamma_{0}}\rightarrow1$ while in the 
ultrarelativistic case ($R_{0}>>1$) \ $K_{\Gamma_{0}}\simeq\left(  2/3\right)
^{1/2}/R_{0}$.

From  Eq. (\ref{R22}) it is clear that $k_{\lim}=\left\vert A_{0}\right\vert
K_{\Gamma_{0}}<<1$ and the instability increment $\chi$ reaches its maximum
$\ \ \chi_{m}=\left\vert A_{0}\right\vert ^{2}K_{\Gamma_{0}}^{2}/4$\ for
the transverse wave vector $k_{\perp m}=2^{-1/2}\left\vert
A_{0}\right\vert K_{\Gamma_{0}}$, corresponding to the characteristic spatial
scale of filaments $\Lambda_{\perp}=\pi/k_{\perp m}$. \ It is worth noting that by
increasing the value of $R_{0}$ the spatial scale of filaments increases
as well. The "critical" power of the EM field carried by the filaments could
be estimated as $P_{c}=\pi$ $\left\vert A_{0}\right\vert ^{2}%
\Lambda_{\perp}^{2}/4\simeq7.8K_{\Gamma_{0}}^{-2}$ which, as it is evident, does not depend on
the pump strength and increases with $R_{0}$. The mentioned critical
power is expressed as $P_{cd}=\left(  m_{e}^{2}%
c^{5}/4e^{2}\right)  \left(  \omega/\omega_{e}\right)  ^{2}P_{c}\approx1.7
\times 10^{16}K_{\Gamma_{0}}^{-2}\left(  \omega/\omega_{e}\right)  ^{2}$erg/sec.
\ A physically reasonable interval of allowed densities of the degenerate electron
plasma is within $(10^{24}-10^{34})cm^{-3}$ $\left[  R_{0}%
\sim1.2-25\right]  $, leading to the following value of the power $P_{cd}\approx\left(
17\times10^{-3}\div10.6\right)  \left(  \omega/\omega_{e}\right)  ^{2}\times 10^{19}$erg/sec.

\begin{figure}
\resizebox{1\textwidth}{!}{%
  \includegraphics{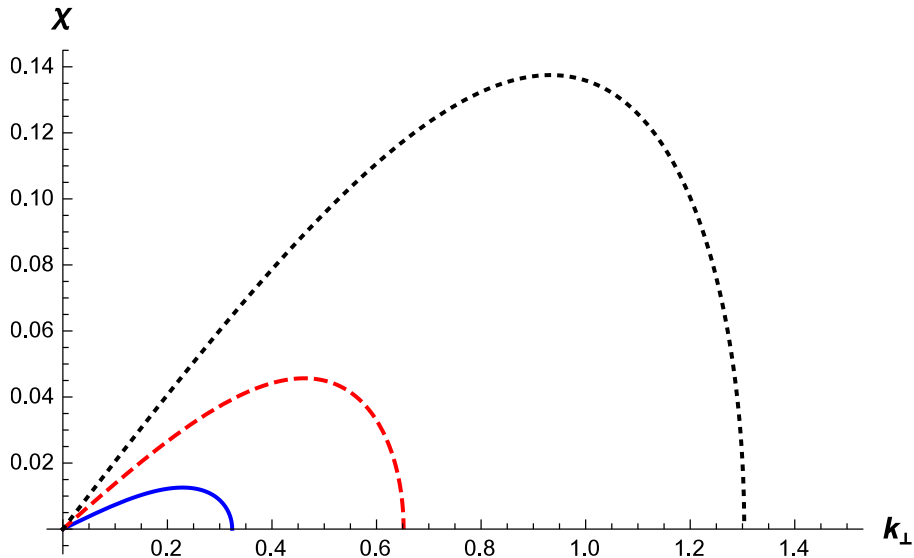}
  }
\caption{Dependence of instability increment $\chi$ on  $k_{\perp}$ for
			different values of $A_{0}$: $A_{0}=0.5$ - solid line, $A_{0}=1$ - dashed line, $A_{0}=2$ -dotted line,}
\label{Figure 2}       
\end{figure}

\begin{figure}
\resizebox{1\textwidth}{!}{%
  \includegraphics{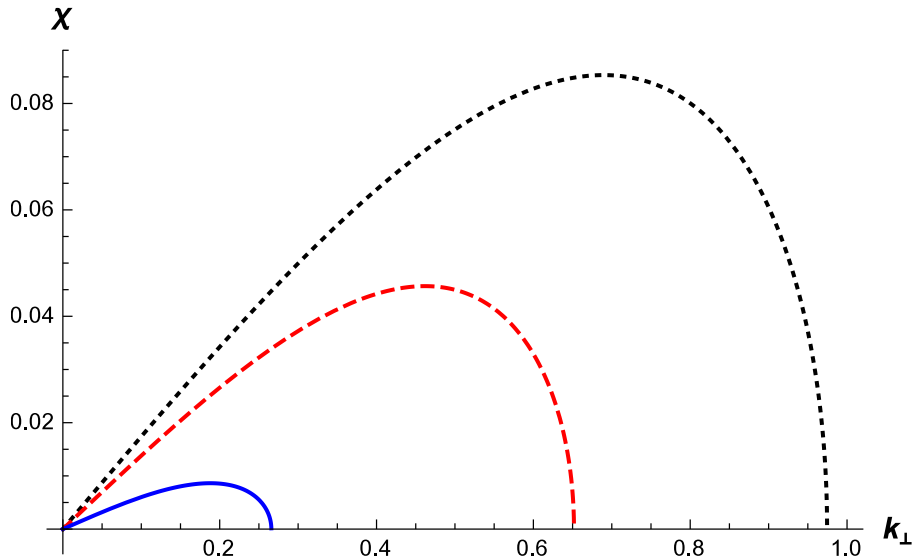}
  }
\caption{Dependence of instability increment $\chi$ on  $k_{\perp}$ for
		different values of $R_{0}$: $A_{0}=3$ - solid line, $A_{0}=1$ - dashed line, $A_{0}=0.2$ - dotted line.}
\label{Figure 3}       
\end{figure}

To calculate the increment of FI for the arbitrary strength of EM
pump the numerical analyses of Eqs. (\ref{R14}),
(\ref{R19})-(\ref{R21}) has been performed. In Fig.\ref{Figure 2} \ for $R_{0}=1$ we show the dependence of $\chi$ on $k_{\perp}$ for different values of $\left\vert A_{0}\right\vert $. It is clear that the maximum of
the instability increment and the range of $k_{\perp}$ over which an instability occurs
$\left(  0<k_{\perp}<k_{\lim}\right)  $ \ increase with increasing values of $\left\vert
A_{0}\right\vert $. Such a behaviour is
valid for the arbitrary strength of degeneracy parameter $R_{0}$. Unlike the regime of 
weak amplitudes $\left(
\left\vert A_{0}\right\vert <<1\right)$ in ultrarelativistic case
$\left\vert A_{0}\right\vert \geq1$ for the characteristic wave vectors of unstable
modes we have $k_{m}\left(  k_{\lim}\right)  \geq1$. In Fig.\ref{Figure 3} we demonstrate dependence
of $\chi$ on $k_{\perp}$ when $\left\vert A_{0}\right\vert =1$ for the different
level of degeneracy $R_{0}$. With increase of the degeneracy level the growth rates
and $k_{\lim}$ decrease.

It is worth noting that since we deal with X-ray sources
the wave amplitude $\left\vert A_{0}\right\vert $ is not
arbitrarily large in the framework of the developed approach. In particular, $\left\vert A_{0}\right\vert $
(in units $eA_{0}/m_{e}c^{2}$) can be given as $\left\vert A_{0}%
\right\vert ^{2}=3.65\times10^{-12}\times I \times
\lambda^2\left[  \mu\right]$, where $I$ is the EM wave intensity
measured in $erg/(sec\; cm^{2})$ and the wave length $\lambda$ is in micrometers. For
$\left\vert A_{0}\right\vert =10$ and $\lambda=0.1nm$ ($\hbar\omega=12.4KeV$)
\ $I\simeq\allowbreak2.\,\allowbreak7\times10^{35}erg/(sec\; cm^{2})$ while electric field of the wave is $E\simeq3\times10^{15}V/cm$ which is by
order of magnitude smaller than the Schwinger limit $E_{S}=m_{e}^{2}c^{3}%
/e\hbar=1.3\times10^{16}V/cm$. Above this limit the EM field is expected to become
nonlinear and the considered approach fails. In particular, under these circumstances, the process of electron-positron pair creation can take place.

\section{Conclusion}
We have shown that the powerful EM beam propagating in the underdense
degenerate plasma undergoes the FI. This process
takes place regardless of degeneracy, implying that the beam power is
larger than the critical power of the appeared structures. The dynamics of formation of filaments in the nonlinear stage and the  subsequent complex behavior should be
investigated by means of numerical simulations of Eqs.(\ref{R17})-(\ref{R18}) and is
beyond of the intended scope of the current paper.

The study of the FI driven by extremely strong EM pulses is significant for understanding dynamics of electromagnetic emission originating from a certain class of astrophysical objects. In particular, it has been widely accepted that $X$-ray emission might appear from accreting white dwarfs (WD) \citep{wd}. In this scenario, the mentioned object accretes material from a companion star, resulting in the hitting process of plasma flow on a star's surface and by means of the Bremsstrahlung mechanism the particles decelerate, leading to generation of $X$-ray radiation. On the other hand, the WDs are composed of highly degenerate electrons and the study of their interaction with the induced $X$-rays might be very promising. The similar process might take place also in neutron stars (NSs) loaded by accretion disk generating hard $X$-rays \citep{Shapiro}. Interiors of NSs mainly consist of neutrons with approximately $1\%$ of electrons and protons \citep{Shapiro}, which are also in highly degenerate state, therefore, the FI might be of high significance. Another interesting class of objects where the aforementioned process might develop are the gamma ray bursters (GRB). It is thought that the high energy radiation of GRB might be generated during a supernova explosion of relatively massive stars, which after they collapse form NSs \citep{carroll}. Since this manuscript was a first attempt of this kind the application of the developed model to the mentioned astrophysical objects is beyond the scope of the paper.

The filamentation instability might be interesting in the case of interaction of superstrong electromagnetic radiation with laboratory plasmas \citep{mourou}.  Such plasmas imbedded in the field of the superstrong laser radiation, with intensities of the order of $10^{21-23}$W cm$^{-2}$ can exhibit various interesting phenomena including self-focusing and FI. The most of the super powerful lasers currently are operating at the wavelength $\lambda\sim 1$ $\mu m$ ($\hbar\omega\sim 1.2$eV). Since the density of degenerate plasma by several orders of magnitude exceeds the critical density for such micron wavelength laser pulses the plasma is opaque. We would like to emphasise that in the standard scheme of the fast ignition model \citep{tabak} plasma is supposed to be compressed at temperatures as low as possible, while by means of a powerful ultra short beam the ignition should occur. However, a compressed target can be in degenerate states \citep{eliezer}. Such plasma states can be transparent just for $X$-ray lasers. Recent achievements in the $X$-ray free-electron laser technology made it possible to achieve intensities above $10^{20}$W cm$^{-2}$ at $9.9$keV ($\lambda\sim 1.3\times 10^{-4}\mu$m, hard $X-rays$) \citep{laser}. This achievement gives us a hope that the increase of $X$-ray laser pulse intensities are feasible. Effects like FI (self-focusing) can be significant in producing $X$-rays with small spot sizes in the processes of interaction with highly compressed degenerate plasmas.

 \section*{Acknowledgments}
The research was supported by the Shota Rustaveli National Science Foundation of Georgia Grant project No. DI-2016-14. VIB acknowledges the partial support from Shota Rustaveli National Science Foundation of Georgia Grant project No. FR17-391.

%
%

\end{document}